\documentclass[english,pre,floats,superscriptaddress,usenames]{revtex4}

\usepackage{cleveref}
\usepackage{array}
\usepackage{booktabs}
\usepackage{listings}
\usepackage{lmodern}
\usepackage{mathpazo}
\usepackage{microtype}

\usepackage{bm} 
\usepackage{amsmath}
\usepackage{amsbsy}

\usepackage{graphicx}
\usepackage{dcolumn}
\usepackage{bm}
\usepackage{hyperref}
\usepackage{soul}
\usepackage[usenames]{color}
\definecolor{Redcolor1}{RGB}{254,0,0}

\newcommand{\bea}{\begin{eqnarray}}
\newcommand{\eea}{\end{eqnarray}}
\newcommand{\be}{\begin{equation}}
\newcommand{\ee}{\end{equation}}

\newcommand{\rr}{{\bf r}}

\newcommand{\FF}{{\bf F}}

\newcommand{\gs}{{\bf g}}
\newcommand{\fs}{{\bf f}}

\newcommand{\vv}{{\bf v}}
\newcommand{\cc}{{\bf c}}
\newcommand{\uu}{{\bf u}}

\newcommand{\Fi}{\Phi}

\graphicspath{{figures/}{}}

\begin{document}

\title{Electrokinetic Lattice Boltzmann solver coupled to Molecular Dynamics: application to polymer translocation}

\author{Adwait V. Datar}
\affiliation{Institute for Computational Physics, Universit\"at Stuttgart, Allmandring 3, 70569 Stuttgart, Germany}

\author{Maria Fyta}
\affiliation{Institute for Computational Physics, Universit\"at Stuttgart, Allmandring 3, 70569 Stuttgart, Germany}

\author{Umberto Marini Bettolo Marconi}
\email{umberto.marinibettolo@unicam.it}
\affiliation{Scuola di Scienze e Tecnologie, Universit\`a di Camerino,
Via Madonna delle Carceri, 62032 Camerino, Italy}

\author{Simone Melchionna}
\affiliation{ISC-CNR, Istituto Sistemi Complessi, Dipartimento di Fisica, Universit\`a Sapienza, P.le A. Moro 2, 00185 Rome, Italy}


\begin{abstract}

We develop a theoretical and  computational approach 
to deal with systems that involve a disparate range of spatio-temporal scales, such as those comprised of  colloidal particles or polymers moving in a fluidic molecular environment.   
Our approach is based on a multiscale modeling  
that combines the slow  dynamics of the large particles
with the fast dynamics of the solvent into a unique framework. The former is numerically solved via Molecular Dynamics and the latter via a multi-component Lattice Boltzmann. The two techniques are coupled together to allow for a seamless exchange of information between the descriptions. Being based on a kinetic  multi-component description of the fluid species, the scheme is flexible in modeling charge flow within complex geometries and ranging from large to vanishing salt concentration. The  details of the scheme are presented and the method is applied to the problem of translocation of a charged polymer through a nanopores. In the end, we discuss the advantages and complexities of the approach.
\end{abstract}

\maketitle


\section{\label{sec:intro} INTRODUCTION}

In the last years, considerable attention has been devoted to the study of non-equilibrium states, such as flowing matter
under the influence of applied external fields,  temperature and/or chemical gradients, pressure differences etc.
The continuum or hydrodynamic description
well describes the movement of fluids in the limit of sufficiently small gradients of density, velocity and temperature, but is inadequate when the variations
of density, velocity  or temperature occur on a length-scale comparable  with the mean free path, 
a situation occurring when molecular fluids flow in confined geometries \cite{eijkel2005nanofluidics}, such as pores \cite{bocquet2010nanofluidics}, nanotubes \cite{mattia2008review}, or when macro-particles move in a molecular solvent \cite{dunweg2009lattice}.  In the latter case, the system is characterized by the presence of disparate length and time scales which cannot be efficiently captured neither by a continuum description
nor by a Lagrangian approach for the time evolution of the individual constituents. 
The dynamics of a  solution of colloidal particles, red blood cells, polymers, proteins or DNA molecules in aqueous solution is characterized by a large difference in size between the fluid molecules of the host fluid and the suspended particles. The Molecular Dynamics (MD) approach  would hardly be efficient due to the differences in the characteristic  times of the dynamics of the two components which would render the approach very time consuming,  whereas the Eulerian approach based on continuous hydrodynamic fields such as the densities and momentum and energy fluxes of each component would neglect the discrete nature of the large component. 

Kinetic theory connects the atomistic and continuum levels and can  treat with success
solutions whose components have similar  physical properties, but when this condition does not hold, as in  the case of a polymer (and other large biological molecules), the method is practically inapplicable. In fact, progress with  the  kinetic approach is guaranteed only when
interparticle correlations are weak and short lived, whereas the motion of each monomer is highly correlated to the motion of other monomers.

Recently, a multiscale approach has been proposed to 
treat  solutions of constituents of disparate sizes and the suitability of the approach to neutral solutions of colloids, polymers and in the context of polymer translocation in nanopores has been described \cite{ahlrichs1999simulation,fyta2006multiscale,dunweg2009lattice}.
The method combines the Molecular Dynamics (MD) method for the suspended particles
with the Lattice Boltzmann (LB) method for the fluid solvent. The advantage of
the LB-MD approach is that hydrodynamic interactions between the solute particles and the
solvent are handled explicitly, and constitute computationally  efficient tool.

When styding charged solutes in an electrolytic solution, one could rely on the assumption of local neutrality, a very common simplification to simulate physical systems justified by the need to reduce the computational effort. Nonetheless, in many cases, electrostatics is essential and an efficient calculation of the long range electrostatic interactions is desirable. The situation is even more complicated under flow conditions, whereby simplified solutions based on local or global equilibrium assumptions break down.

Electrokinetics involves Coulomb interactions between charged particles, as well as the motion of these particles within a flow field \cite{masliyah2006electrokinetic}. 
Its effects are  of high interest in Microfluidics \cite{li2004electrokinetics} or Soft Matter and Biophysics. These areas involve for example, the behavior of colloids \cite{schmitz2013computer},  biomolecules in a flow \cite{fogolari2002poisson,lu2008recent}, active swimmers \cite{de2016lattice}, various types of solutions \cite{fischer2008salt}, ion transport \cite{pennathur2005electrokinetic}, etc. 
Electrokinetics is also responsible for electrophoretic and electroosmotic phenomena
in micro and nanoconfinement \cite{sparreboom2010transport,westermeier2016electrophoresis,spanner1975electroosmotic}. 

%

In order to include electrokinetic effects, one can account for different levels of description of the forces occurring in the system: the electrostatic potential asiring from the Poisson equation, the diffusion and advection of the charges described by the Nernst-Planck equation, and the velocity for the fluid flow described by the Navier-Stokes equation. Given the description of preference, a discrete grid is often taken in order to solve these coupled equations numerically. 
In the past, various simulation approaches have been developed for solving electrokinetics  problems. One typical example is the Poisson-Boltzmann-Nernst-Planck model \cite{zheng2011poisson}, which couples the Nernst-Planck equations with the Boltzmann distributions of ion concentrations. Mesoscopic approaches are commonly based on the LB method including an electrohydrodynamic coupling, and well capture electrokinetic phenomena in complex systems like colloidal suspensions  \cite{pagonabarraga2005mesoscopic,kuron2016}. Electroviscous transport phenomena can also be captured by this route  \cite{warren1997electroviscous}.  A microscopic self-consistent  coupling of the relevant equations of kinetic theory, classical density functional theory, and LB are approaches capable of studying electro-osmotic flows under nanoconfinement or the modulation of the ionic current to  DNA docking in a nanopore  \cite{marini2012charge,melchionna2011electro,chinappi2014modulation}. Other numerical algorithms coupling the Poisson-Nernst-Planck equation for the electrostatic potential with the classical equilibrium density functional theory have been used  to model ion channels and extract the ion flux through ion-selective pores  \cite{gillespie2002coupling,rosenfeld1993free}. On another level, MD is also quite efficient in investigating ion gating in water filled channels \cite{leung2008ion}. A combination of MD simulations with a mesoscopic description and a coarse-graining scheme can evaluate the ion dynamics in clay interlayers 
\cite{rotenberg2007multiscale}. 

We extend here a previous approach to representing polymers embedded in a single neutral fluid, inclusive of hydrodynamic interactions \cite{ahlrichs1999simulation,fyta2006multiscale}. 
The  numerical scheme also takes into account the electrostatic and viscous interactions \cite{marconi2011dynamics,marini2012charge}, but not the presence of charged solutes. In this paper, we describe the joint methodology so to treat an electrolytic solution and charged or neutral suspended particles or molecules in full generality.
To this purpose, the main strengths of our approach are: the possibility to consider multiple fluid components on the same footing, in particular in the limit of vanishing concentrations where otherwise a Lagrangian approach would give rise to complicated averaging procedures, the inclusion of chemical specificity for the suspended particles, the joint solution of electrostatics for the solvent and solute species, and finally the intrinsic computational efficiency of the method, that is particularly prone to extensive parallelism.



The paper is organized as follows: in Section \ref{sec:method}, the multiscale approach is introduced together with the main ingredients of the electrokinetics scheme and the coupling to the Molecular Dynamics methods. In Section \ref{sec:benchm} the multiscale scheme is applied to the process of polymer translocation through narrow pores and in Section \ref{sec:summ} the concluding remarks are given.

\section{Methodology\label{sec:method}}

The multiscale scheme presented in the following is based on solving numerically the fluid equations via the LB-MD methodology. Therefore, the LB solver describes multiple fluid components in the Eulerian framework and the suspended particles in the Lagrangian framework. The reason for this distinction is ultimately due to the need to follow the identity of each suspended particle, as we shall see hereafter.

\subsection{Continuum description of the solution}

At first, we intend to describe how a system composed of a solvent and an electrolyte can be described by taking into account both the electrostatic and viscous interactions and the transport of each species.
To this purpose, we consider a ternary charged mixture comprised of a neutral species, a positively and negatively charged species. We consider the fluid as composed of point-like particles and neglect correlations stemming from excluded volume interactions. Similarly, the charged species carry point-like charges (the ions).
Each species is denoted by $\alpha$, has mass $m_\alpha$, and two ionic components carry charges $z_\alpha e$, with $e$ being the proton charge. The index $\alpha=0$ identifies the neutral solvent, while $\alpha \pm 1$ identifies the two oppositely charged ionic species interacting through a uniform medium of constant dielectric permittivity, $\epsilon$. The solution accommodates an arbitrary number of molecules, with the i-th particles having position $\rr_i$, velocity $\vv_i$ and valence $z_i$.

Following the Boltzmann description, the state of each fluid species is determined by the distribution function $f^\alpha(\rr,\vv,t)$ being proportional to the probability of finding at time $t$ the fluid species $\alpha$ at position $\mathbf{r}$ and moving with velocity $\mathbf{v}$. The description is mesoscopic, neglects the identity of each fluid molecule and is based on a statistical, coarse-grained representation of the solvent. 

Starting from the distribution function, we derive the macroscopic variables. In particular, $$n^\alpha(\rr,t) = \int dv f^\alpha(\rr,\vv,t) $$ is the zeroth moment and 
$$n^\alpha \uu^\alpha(\rr.t) = \int d \vv \vv f^\alpha(\rr,\vv,t) $$
 is the first moment of $f^\alpha$, respectively. In addition, 
 $\rho^{\alpha} = m^{\alpha}n^{\alpha}$
  is the mass density and, for the multicomponent system, we define the current of a given species relative to the center of mass motion as
\begin{equation}
\mathbf{J}^{\alpha}(\rr,t) = \rho^{\alpha}[\uu^{\alpha}(\rr,t) - \uu(\rr,t)]
\end{equation}
where the baricentric velocity is
$\uu = \frac{\sum_{\alpha}\rho^{\alpha}\uu^{\alpha}}{\sum_{\alpha}\rho^{\alpha}}$.
In addition, $n=\sum_\alpha n^\alpha$ is the total number density. 

The electrostatic potential $\psi(\rr,t)$ generated by the charge distribution 
\begin{equation}
\rho_e(\rr,t) = e[n^+(\rr,t) - n^-(\rr,t) + N(\rr,t)]\, ,
\end{equation}
 where $N=\sum_i z_i \delta(\rr_i - \rr)$ is the charge density of suspended particles, and by the surface charge density $\Sigma(\mathbf{r})$ associated with the presence of charges on the walls, is obtained by solving Poisson equation for electrostatics
\begin{equation}
\nabla^2 \psi(\rr) = - \frac{\rho_e(\rr)}{\epsilon}
\label{eq:poisson}
\end{equation}
and by imposing Dirichlet or Neumann boundary conditions. For insulating confining walls, this reads $-\nabla \psi \cdot \hat n = \Sigma / \epsilon$, where $\hat n$ is the unit vector normal to the surface. 


The Boltzmann equation for the multicomponent solution reads
\begin{eqnarray}
\left[
\frac{\partial}{\partial t} + 
\vv\cdot\nabla +
\frac{\Fi^{\alpha}}{m^\alpha} \cdot \frac{\partial}{\partial \vv}
\right] f^{\alpha} =  -\omega_{visc} (f^{\alpha} - f_{eq}^{\alpha})
\label{eq:LBmulti}
\end{eqnarray}
where the right hand side is a collisional operator that takes into account viscous  forces stemming from fluid-fluid interactions,
and has the Bhatnagar-Gross-Krook (BGK) form \cite{bhatnagar1954model} that relaxes the distribution function towards the Maxwellian equilibrium distribution.
In the left hand side, the total force $\Fi^{\alpha}$ acting of the fluid species $\alpha$ is the sum of different terms:

\begin{equation}
\Fi^\alpha(\rr) = \Fi^{\alpha}_{ext}(\rr)  - ez^\alpha \nabla \psi(\rr)+ \Fi^{\alpha}_{drag}(\rr) + \sum_{i=1}^N \gs^\alpha_i \theta_i(\rr) 
\label{totalforcefluid}
\end{equation}
that is, the sum of forces acting on the species $\alpha$ of the  fluid, where
$\Fi^{\alpha}_{ext}$ is the external force,  $ - ez^\alpha \nabla \psi$ is the electrostatic force. The term
\begin{equation}
\Fi^{\alpha}_{drag}(\rr) = - m^\alpha \omega^{\alpha}_{drag} \sum_\beta \frac{n^\beta(\rr)}{n} (\uu^\alpha(\rr)-\uu^\beta(\rr))
\label{Fidef}
\end{equation}
 is the drag force exerted on species $\alpha$, { that results in a cross-diffusion coefficient $D^\alpha = \frac{v_T^2}{\omega^{\alpha}_{drag}}$, where $\omega^\alpha_{drag}$ is a relaxation frequency
 related to the collision
  rate between the ions and the solvent and $v_T$ the  thermal velocity. 
Equation~\eqref{Fidef} corresponds to the following  ionic mobilities 
 $\lambda^\pm = \frac{D^\pm}{k_B T}=\frac{1}{m^\pm\, \omega^\pm_{drag}}$.
Finally, $\sum_{i=1}^N \gs^\alpha_i \theta_i(\rr)$ is the drag force arising from the $N$ particles in  solution, as detailed in the next section, and $\theta_i(\rr)$ is the (adimensional) particle occupation number, being equal to $1$ when the ith particle is sitting in a elementary cube of volume $\sigma^3$ around position $\rr_i$, and zero otherwise.
The function $\theta_i(\rr)$ is introduced in order to bridge the continuum Boltzmann equation with the 
MD description, or in other words  to combine the Eulerian (field based description) with the Lagrangian (particle based description) worlds into the same framework.}

{ 
Representing the microscopic motion of the suspended molecule requires considering the spontaneous thermal fluctuations of the solute, fluctuations that do not subside at the nanoscopic scale and are a major component among the forces that regulate translocation. By the same token, taking into account fluctuations for the solution can be considered via a stochastic version of the kinetic equation. To this purpose and following Landau-Lifshitz fluctuating hydrodynamics, a source of random fluctuations is added on the momentum-flux tensor. The specific form of the fluctuations is described in the following.
}

\subsection{Particle description of the suspended particles}

For the MD part, the Langevin equation describes the dynamics of the $i^{th}$ particle of the suspended particles, as for the i-th bead of a polymeric chain: 
\begin{subequations}
\begin{align}
\frac{d}{dt}\mathbf{\dot{R}}_i & = \frac{1}{m_i}\mathbf{p}_i\\
M_i\frac{d}{dt}\mathbf{\dot{V}}_i & = 
-\sum_\alpha \gs^{\alpha}_i 
+ \FF_i + \fs_i
\end{align}
\end{subequations}
where $\mathbf{R}_i$, $\mathbf{V}_i$ and $M_i$ are the particle position, velocity and mass, respectively, $\gs^\alpha_i = - \gamma \left[\uu^\alpha(\rr_i) - \mathbf{V}_i\right]$
is the sum of drag forces arising from all fluid-particle interactions, and 
$\fs_i$ is the thermal noise acting on the particle obeying
the fluctuation dissipation theorem:
\begin{subequations}
\begin{align}
\langle f_{i\alpha}\rangle & = 0 \\
\langle f_{i\alpha}(t)f_{j\beta}(t')\rangle & = 2\gamma_iK_BT\delta_{ij}\delta_{\alpha\beta}(t-t') 
\end{align}
\end{subequations}
$\FF_i = - \sum_j \nabla_i U + q_i \nabla \psi $ is the conservative force containing the inter-particle (with $U(\mathbf{R}_i)=\sum_{ij} u (R_{ij})$) and electrostatic interactions. The particle-particle interactions contain excluded volume, in the form of the truncated Lennard-Jones or Weeks-Chandler-Anderson potential \cite{weeks1971role},
and bonding terms among beads of the chain, reading:
\begin{subequations}
\begin{align}
u(R_{ij}) & = 
4\epsilon\left[
\left(\frac{\sigma}{R_{ij}}\right)^{12} - \left(\frac{\sigma}{R_{ij}}\right)^6
\right]  \textnormal{~for} \: R_{ij}<2^{\frac{1}{6}}\sigma,~ |i-j|>1\\
& = 0  \textnormal{~~~~~~~~~~~~~~~~~~~~~~~~~~~~~~~~~for} \: R_{ij}<2^{\frac{1}{6}}\sigma,~|i-j|>1\\
& = \frac{k(R_{ij}-l_{\textnormal{bond}})^2}{2} \textnormal{~~~~~~~~~~~~~~~~for~}|i-j|\leq 1
\label{polymermodel}
\end{align}
\end{subequations}
where $R_{ij} = |\mathbf{R}_i - \mathbf{R}_j|$ is the distance between the ith and jth particles, 
$\epsilon$ and $\sigma$ are Lennard Jones parameters, $k$ the bond force constant and
$l_{\textnormal{bond}}$ the bond length between two adjacent beads. Often the molecular part involves different time scales than the mesoscopic LB part. Accordingly, our scheme copes with these variations by accounting for different timesteps for the LB and MD part, $\Delta t$ and $\Delta t_{MD} = \Delta t/n$, respectively, and by employing a temporal subcycling of $n$ steps for the MD solver.

\subsection{Discretization procedure}

The numerical solution of the coupled equations for the distribution functions $f^\alpha$  is obtained by employing  the LB method, as presented in Ref. \cite{chen1998lattice}. The numerical procedure is a modification of the  method used in fluid dynamics applications by the addition of the solute that is treated as a suspended body in the ternary fluid.


In a nutshell, the LB method \cite{higuera1989lattice,succi2001lattice,chen1998lattice} comprises the following steps.
We discretize eq. \eqref{eq:LBmulti} by choosing a spatio-temporal discrete grid, with 
a unit timestep ($\Delta t=1$), and unit lattice spacing ($\Delta x=1$).
Thus, the position coordinate is discretized by introducing a Cartesian mesh whose lattice points 
are separated by $\Delta x$. The continuous velocity is also discretized and takes  only
$Q$ possible values, $\{\cc_p\}$ with $p=0,Q-1$. 
The discrete velocities are chosen as to connect neighboring spatial mesh points.
For the present three-dimensional study, we use a 19-speed lattice, the so-called D3Q19 scheme, consisting of one speed $\cc_p= 0$ for particles at rest on a mesh node, 6 discrete velocities with $|\cc_{1-6}|=1$ and pointing towards the first mesh neighbors, and $12$ velocities with $|\cc_{7-18}|=\sqrt 2$ pointing towards the second mesh neighbors.
The  continuous phase space distribution functions $f^\alpha$ 
are replaced by the arrays $f^\alpha(\rr,\vv,t)\to f^\alpha_p(\rr,t)$.
 Their  equilibrium expressions can be written in the following form:
\begin{equation}
f_{p}^{\alpha,eq} = w_{p}\left[n^\alpha+n^\alpha \frac{\delta \uu^\alpha\cdot\cc_p} {v_T^2}+
n^\alpha \frac{ (\delta \uu^\alpha\cdot\cc_{p})^2-v_{T}^{2} (\delta u^\alpha)^2)}{2v_{T}^4} \right]
\label{eq:kernelf}
\end{equation}
where $\delta \uu^\alpha=\uu^\alpha-\uu$.
The discretized form of the eqs. \eqref{eq:LBmulti} corresponds to  the following propagation scheme:
\begin{equation}
{f}^\alpha_{p}(\rr+\cc_{p},t+1) = {f}^\alpha_{p}(\rr,t) + \Omega^\alpha_{p}(\rr,t)
\end{equation}
where the term 
{ 
\begin{equation}
\Omega^\alpha_{p}(\rr,t) = -\omega_{visc} (f^\alpha_{p}(\rr,t)-f_{p}^{\alpha,eq}(\rr,t) )+S^\alpha_{p}(\rr,t) + \Delta^\alpha_p(\rr,t)
\label{eq:kernelf2}
\end{equation}
is the sum of the BGK viscous relaxation contribution plus 
contribution accounting for drag, external, electrostatic via the $S^\alpha_p$ term, and fluctuating forces via the $\Delta^\alpha_p$. In the above, we set the fluid masses $m^\alpha=1$ for simplicity.
}
The contribution stemming from drag, external and electrostatic forces is written as:
\begin{equation}
S^\alpha_{p}(\rr) = w_{p} n^\alpha(\rr) 
\Bigl[ \frac{\Fi^\alpha(\rr) \cdot \cc_p}{v_T^2} + 
\frac{ (\cc_{p}\cdot \uu^\alpha(\rr))(\cc_{p}\cdot \Fi^\alpha) -v_{T}^{2}\Fi^\alpha(\rr) \cdot \uu^\alpha)}{v_{T}^{4}} 
\Bigr]
\label{eq:kernelg}
\end{equation}
where $\Fi^\alpha(\rr)$ is the discretized form of the force appearing in Eq.   \eqref{totalforcefluid}
and $w_p$ are Gaussian weights. 
{
Owing to the discrete nature of the forces, these are partially added to the currents in order to achieve higher stability and second-order accuracy of the numerical method, as described in refs \cite{guo2002discrete,melchionna2012stabilized}.

The $\Delta_{p}$ term is local in space and time and acts at the level of the stress tensor and non-hydrodynamic modes. It is constructed via a set of the lattice eigenvectors 
$\{\chi_{k}\}_{k=0,Q}$ orthonormal according to the scalar product $\sum_{m=0}^{Q}w_{m}\chi_{km}\chi_{lm}$. 
In the D3Q19 scheme, the eigenvectors correspond to the kinetic moments: $k=0$ is relative to density, $k=1-3$ to  current, 
$k=4-9$ on the momentum flux tensor, the remaining $k=10-Q$ eigenvectors to 
non-hydrodynamic modes.  The stochastic forcing reads
\begin{equation}
\Delta^\alpha_{p}=\sqrt{m n^\alpha \omega_{visc} (2-\omega_{visc})}\sum_{k=4}^{Q}w_{p}\chi_{kp}{\cal N}_{k}
\end{equation}
where $m$ is the particle mass, equal for all species, and ${\cal N}_{k}$ is a set of $15$ random numbers with zero mean and unit variance \cite{dunweg2009lattice}. 
Given the fact that the thermal velocity is fixed and derives from the underlying lattice, the thermal mass is chosen so to obtain the thermal fluctuations according to $k_BT=m v^2_T$.
The stochastic forcing produces consistent fluctuations at all spatial scales, 
in particular at short distances where the effect on the translocating molecule is critical. 
}

At every timestep, once the populations $f^\alpha_p$ are known, they are used to compute hydrodynamic moments, entering both the equilibrium, and in sampling the macroscopic evolution.
The species densities are evaluated as \cite{marconi2009kinetic}
\be
n^\alpha(\rr,t)=\sum_p f^\alpha_p(\rr,t)
\ee
and the momentum fluxes as:
\be
n^\alpha(\rr,t) {\bf u}^\alpha(\rr,t)=\sum_p \cc_p f^\alpha_p(\rr,t) \,. 
\ee

In order to obtain the  electrostatic contribution to the force at each iteration step we solve 
numerically  the Poisson equation \eqref{eq:poisson} for the electrostatic potential generated by the ions, the surface charges and the polyelectrolyte.
We employ a successive over-relaxation method \cite{press1989numerical} whose
speed of convergence is greatly enhanced by employing a 
Gauss-Seidel checker-board scheme \cite{nicholls1991rapid} in conjunction with Chebychev acceleration \cite{li2005chebyshev}. 

\begin{figure}
 \centering 
 \includegraphics[width=0.6\linewidth]{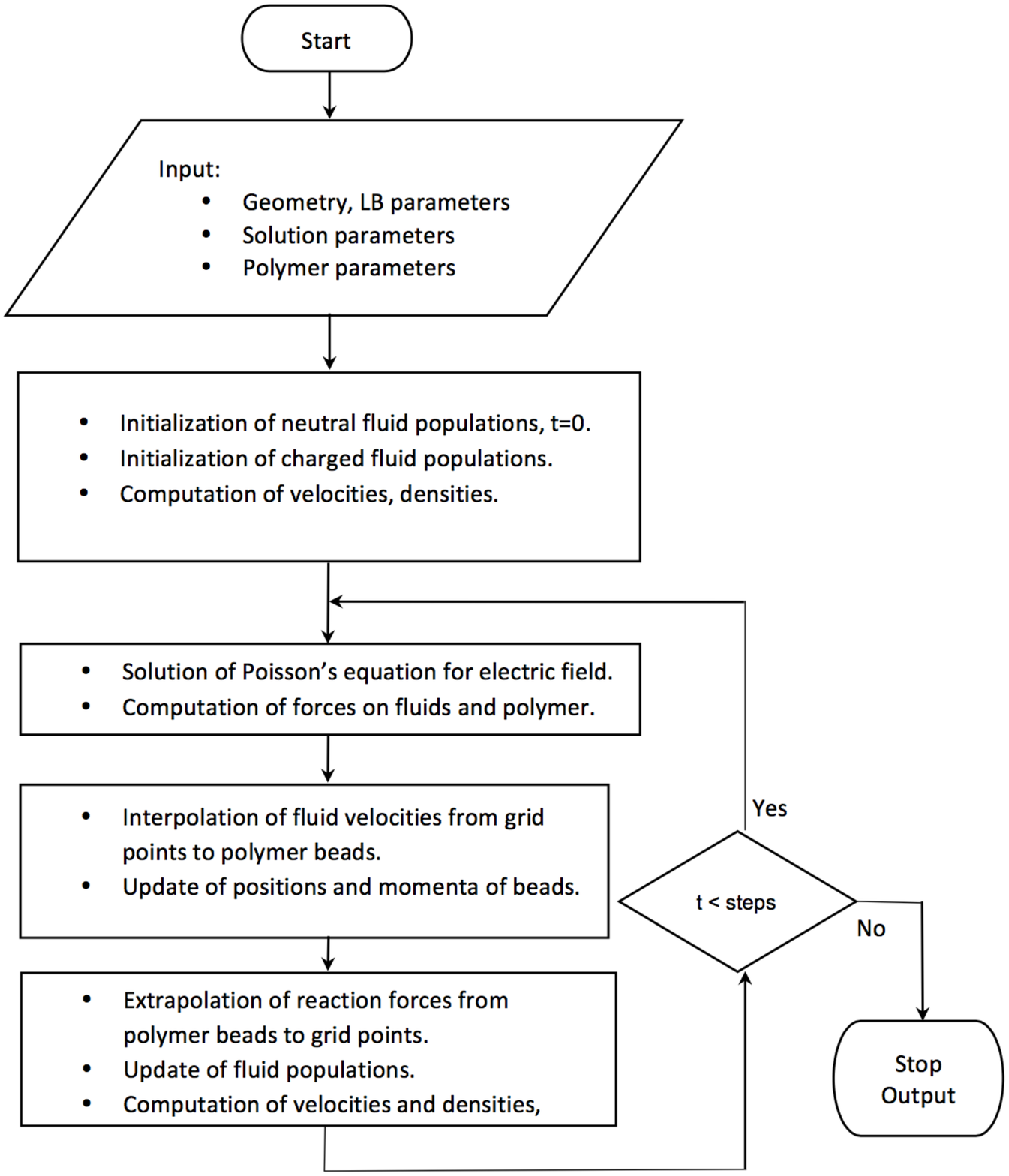}
  \caption{A sketch of the algorithm followed in the multiscale simulations.}
  \label{fig:algo}
\end{figure}

The algorithm followed in our multiscale simulations is sketched in Fig.\ref{fig:algo}. Briefly, the geometry (lattice, boundary conditions), temperature, timesteps and all fluid and molecular parameters are given as input. The simulation starts by initializing the fluid (neutral and charged) populations and computing the corresponding densities and velocities. Poisson's equation is then solved for the electric field giving the forces acting on the fluid populations and the molecular part (polymer). 

The fluid velocity of the nearby grid point is interpolated to the molecular bead at position $\mathbf{R}$. This is obtained by using special interpolation kernels $\tilde{\delta}(\rr,\mathbf{R})$, such as used in other numerical contexts \cite{dunweg2009lattice}. The molecule is then advanced for a fraction of LB timesteps. 
The updated positions and velocities of the beads lead to the forces which are extrapolated back to the LB fluids by the same extrapolation kernels $\tilde{\delta}$. The fluid populations are updated and the LB solver takes over. This tasks are being performed for the given total simulation time (in LB timesteps, $\Delta$t).





\section{Benchmark: Electrophoretic flow of polymers\label{sec:benchm}}

In order to test the validity of the present approach, we consider 
the flexible polyelectrolyte chain described in the previous section
and study its  electrophoretic behaviour, i.e. its response to an externally applied electric field measured by its mobility. Electrophoresis is a standard technique 
to separate and characterize  biomolecules such as proteins and DNA, as well as synthetic polyelectrolytes \cite{slater2009modeling}.
 Under the action of a constant electric field its average drift velocity depends on the field itself, but also on the interactions
 between the polyelectrolyte, the solvent and the ions \cite{grass2008importance}.
 The electrophoretic mobility  curves  in free solution display some universal dependence on the length of a flexible polyelectrolyte \cite{grass2010mesoscale}.
 Short  polymers show an increase in electrophoretic mobility with chain length whereas  for long globular chains  the mobility approaches a constant in the free draining regime. 
 Between these two regimes, counterion and hydrodynamic effects become equally important.
  
In order to validate our model against previous studies, we compute the mobility of a charged linear polymer as a function of its length. 
The polymer is initially prepared in a cubic simulation box of size $256\times 256 \times 256 ~ \Delta x^3$. The electrolytic solution and polyelectrolyte are taken as globally electroneutral, by compensating the charges of the coions to meet the condition. The bead of the polyelectrolyte carries a charge of $0.1 e$. Such value provides stable numerical simulations, which in fact stems from having soft Coulombic forces $ez^\alpha \nabla \psi < kT/\Delta x$, ultimately satistying the Courant-Friedrich-Levy condition for numerical stability \cite{courant1959partial}. In the future, such limitation could be overcome by adopting a multiple timestep approach based on a propagator specifically designed to handle stiff forces \cite{SCHILLER20142586}.
Before activating the electrophoretic force, the initial configuration is initially relaxed by running the system for approximately the Rouse time and at zero electric field. 

\begin{figure}[h]
\centering
\includegraphics[width=0.6\linewidth]{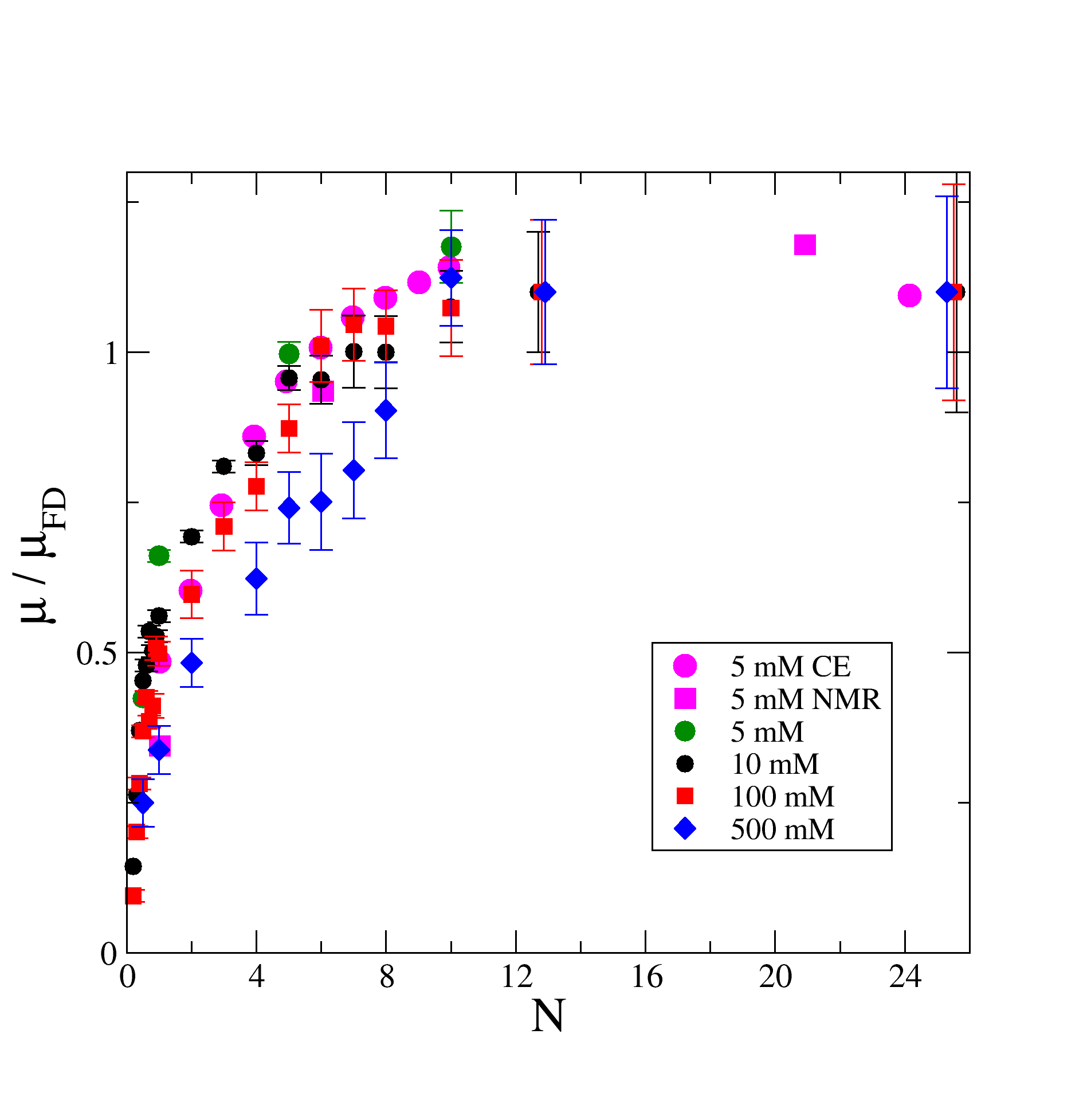}
  \caption{Normalized electrophoretic mobility $\mu / \mu_{FD}$ as a function of the number of repeat units $N$ compared to results from
capillary electrophoresis (CE) and from electrophoretic NMR from \cite{grass2008importance}.}
  \label{fig:benchm_mob}
\end{figure}

The variation of the  electrophoretic mobility with the polymer length is given in Fig.\ref{fig:benchm_mob} for different salt concentrations and is compared to previous experimental results \cite{grass2008importance}. The electrophoretic mobility is normalized with respect to the free-draining mobility $\mu_{FD}$ as obtained for long chains. In this way, the different viscosity of the solvent is taken into account. 

The rescaled electrophoretic mobility shows an increase at large number of polymer units, i.e. the number of beads $N$. The mobility displays a peak as a function of the number of beads. Moreover the larger the electrolyte concentration the lower the mobility. This is justified by the screening of the beads imposed by the mobile ions.
As evident from Fig. 2, our benchmark reproduces the known behavior \cite{grass2008importance} and validates the theoretical approach developed in this work. In this respect, we then move further in using our multiscale scheme to a more complex problem.

\section{Application: polymer translocation \label{sec:transloc}}

The translocation of polymers through narrow pores is a process that has been intensively investigated over the last 20 years \cite{kasianowicz1996characterization}. The passing, i.e. translocation of biomolecules such as DNA, through nanopores shows a high promise for realizing single-molecule experiments and also reading-out biomolecules, such as RNA and DNA for sequencing applications  \cite{zwolak2008colloquium,meller2012nanopores,branton2008potential}. 

Typically, a voltage difference is applied across two chambers filled with a saline solution. The chambers are separated by a thick wall in which a nanopore has been drilled \cite{li2001ion}. One of the two chambers is initially filled with biomolecules, such as DNA or RNA, which are - as the ions of the solution - electrophoretically driven through the nanopore. Electro-hydrodynamics-based models are used to unravel the ion concentrations and electro-osmotic flow in a nanopore for DNA selectivity \cite{luan2013electro}.  From a fundamental point the translocation of polymers is also of high interest because of the underlying interesting physical behavior  \cite{muthukumar2016polymer}. Accordingly, as a benchmark of our method, we take a charged polymer and model translocation events through a narrow charged nanopore. 


\begin{figure}[h]
\centering
\includegraphics[width=0.5\linewidth]{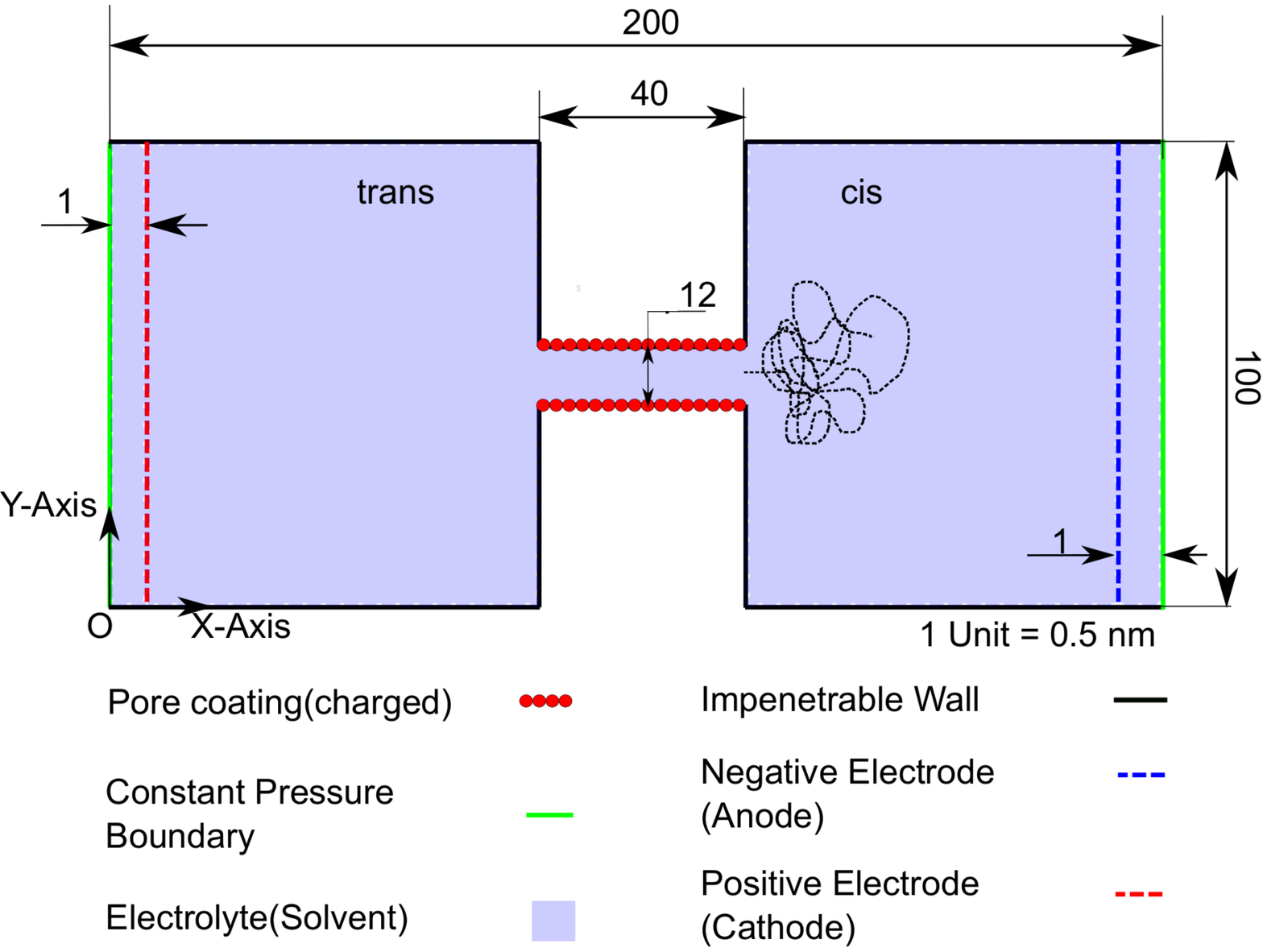}
  \caption{A sketch of the translocation setup. The labels are self-explaining. The polymer shown as a blob in the right is randomly generated and is translocating to the left (from the {\it cis} to the {\it trans} side of the nanopore. All units are given in simulation units for which the mesh spacing corresponds to 0.5 nm, as also shown on the sketch and discussed in the text.}
  \label{fig:setup}
\end{figure}

\subsection{System set-up}

\begin{figure}
\includegraphics[width=0.4\linewidth]{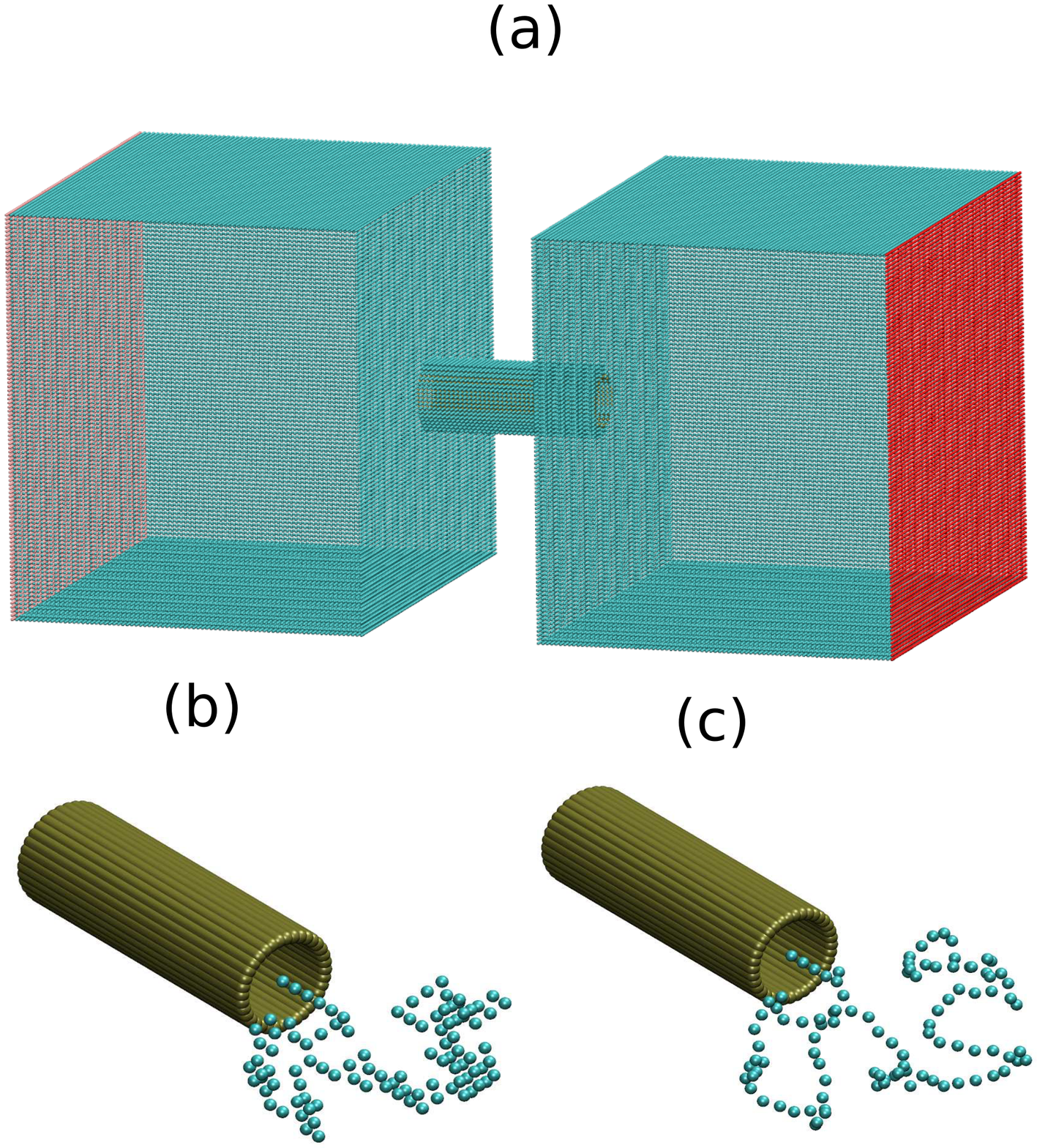}
\caption{(a) A typical setup of the computational box with the nanopore. Panels (b) and (c) depict a close-up view of the nanopore with the polymer before and after the equilibration, respectively.
}
\label{fig:polym}
\end{figure}

A sketch of the setup for the polymer translocation is shown in the Fig. \ref{fig:setup}.
We choose a pore size typical of that in experiments, i.e. a pore diameter of $12$ LB units (6 nm) with a membrane thickness of $40$ units (20 nm). In order to accurately represent the flow in the pore, we chose a fine enough LB grid with a lattice size $\Delta x$ equal to 0.5 nm leading to 12 grid points along the diameter of the pore. The bead-bead bond length is chosen to be 2$\Delta x$, that is, $1 nm$, which is generally used in the literature for simulations of polyelectrolytes \cite{wallin1997monte}. A 3D representation of the lattice for the chambers and the nanopore is depicted in Fig.\ref{fig:polym}(a).
The pore is internally charged by placing an array of charges on the pore wall, with a local negative linear surface of $0.36~ C/nm$.

The polyelectrolyte is relaxed prior to initiating the translocation process. Using a random walk algorithm, a polymer is first generated in the {\it cis} chamber starting from the first bead at the pore entrance (see sketch in that figure and the 'as generated' polymer in Fig.\ref{fig:polym}(b)). We considered a polymer composed of $100$ beads. 

For the LB and the MD part, the same timestep size was taken, $\Delta t=\Delta t_{MD}$. The polyelectrolyte is first equilibrated away of the pore mouth for approximately the Rouse time 
before applying the potential difference as depicted in Fig. \ref{fig:polym} (bottom). A schematic of an equilibrated polymer is shown in Fig.\ref{fig:polym}(c). After the equilibration a potential difference is applied across the chambers and this pulls the charged polymer through the nanopore. In this way, translocation events are monitored. A standard freely jointed chain model for the polymer with $\epsilon = k_BT$ (where $T$ The temperature and $k_B$ the Boltzmann constant) and $\sigma = 1$ nm. As for electrophoresis in bulk, numerical stability requires that we set the charge per bead at $0.1 e$. For an aqueous solution, we took the Bjerrum length for water at $0.7 nm$. The applied electric field was $0.05 ~ k_BT / e \Delta x$ which falls well within the linear regime.

We have simulated translocation events for $100$ randomly initiated polymers, as discussed above. In order to evaluate the dynamics of the translocating polymers, we analyze the times when each beads crosses the midsection of the pore. All simulations led to successful translocation events. About 10 \% of all events were multi-file events during which more than one parts of the polymer chain were found in the pore \cite{li2003dna,merchant2010dna,bernaschi2008quantized}. These events will not be considered further for the analysis presented here. 
In the following, we mainly observe and compare the solvent characteristics for an open pore (without the polymer) and a filled pore (translocated by a polymer) and use the former as a reference. For the filled pore results, we average over all observed (single-file) translocation events.

\subsection{Flow and polymer characteristics}

\begin{figure}[h]
\includegraphics[width=0.5 \linewidth]{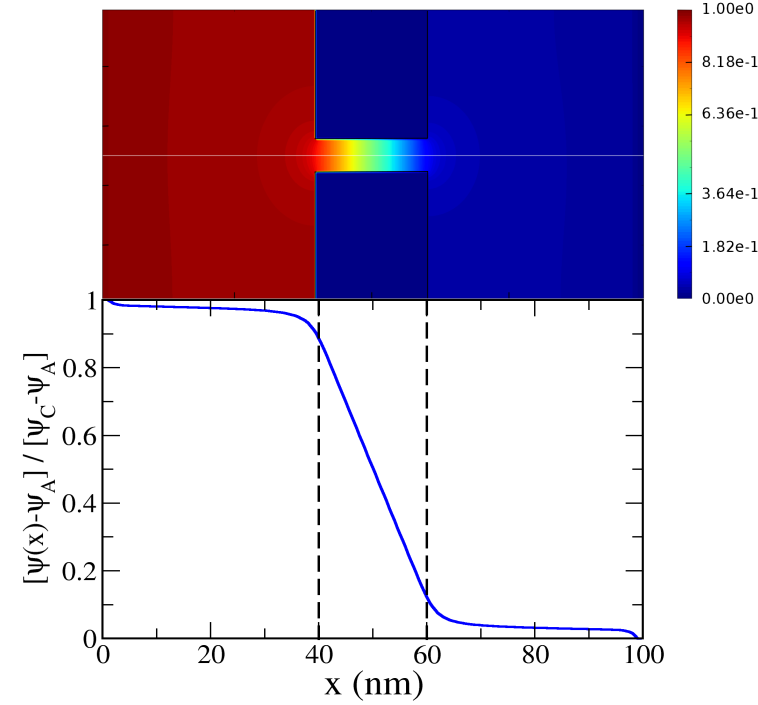}
\caption{Top: Electric potential $[\psi(x)-\psi_A]/[\psi_C-\psi_A]$, where $\psi_C$ and $\psi_A$ refer to the anode and cathode potentials, respectively, for the open pore. Upper panel: contour plot. Lower panel: potential profile along the translocation direction (moving from right to the left). The vertical dashed lines indicate the pore entrance and exit.}
\label{fig:pot_open}
\end{figure}

At a first step, we observe the behavior in the electroosmotic flow (EOF) in the open pore. In Fig.\ref{fig:pot_open},  results are shown for the electric potential across a cut along the translocation direction (from the right to the left). We observe that the electric potential is almost constant in the bulk and drops at the pore. The color differences denote large differences across the chambers. Taking a cross section in the pore perpendicular to the translocation direction shows a constant electric potential denoting that the gradient of this potential is significant only along the translocation direction. This is clear by looking at the variation of the electric potential along the translocation direction (lower panel of Fig.\ref{fig:pot_open}).
It can also be  observed from this figure that the potential drop occurs  only in the pore. Accordingly, the open nanopore system can be thought of as an electrical circuit including a  resistance across which a potential difference is applied. The resulting  linear  variation, which will be discussed in Fig. \ref{fig:conduct} confirms this.
The gradient of the potential, which is the magnitude of the electric field is close to zero in the bulk in both chambers, but is high in the pore. The component of the electric field along the translocation direction is positive in the pore (Fig.\ref{fig:field}), as the electric field 
acting on the cations will point to that direction. 

\begin{figure}[h]
\centering
\includegraphics[width=0.5 \linewidth]{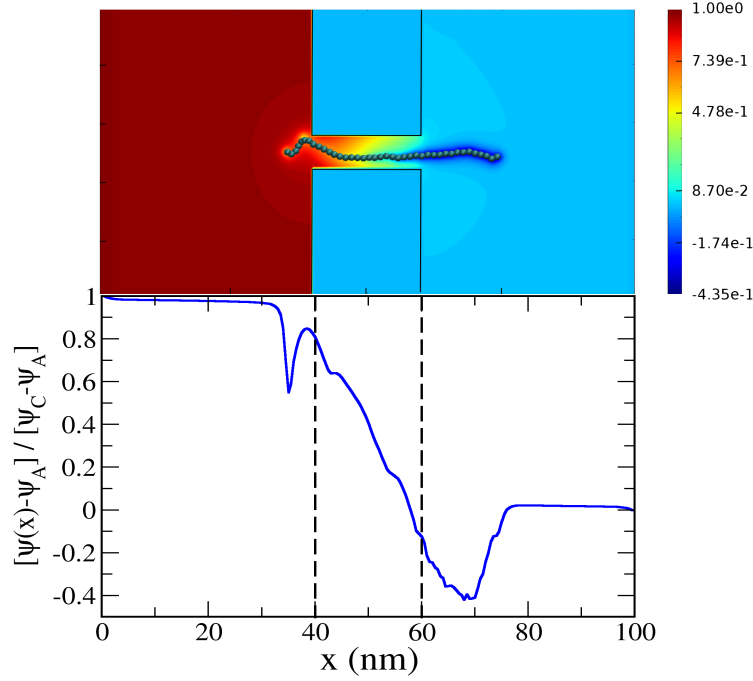}
  \caption{Top: Electric potential for the closed pore, at a given polymer translocation event. The polymer conformation relevant to the  snapshot is also shown. Upper panel: contour plot. Lower panel: potential profile along the translocation direction (moving from right to the left). The vertical dashed lines indicate the pore  entrance and exit.}
  \label{fig:pot_polym}
\end{figure}

The flow behavior in the case of a nanopore which is translocated by a polymer is qualitatively quite different as evident from the electric potential shown in Fig.\ref{fig:pot_polym}. Inspection of the upper panel of this figure, clearly reveals a correlation between the values of the electric potential and the conformation of the translocating polymer. An overall gradient in the electric potential in the pore is also evident. Small fluctuations in the drop of the electric potential along the translocation direction can also be attributed to non-linear polymer configurations. A comparison of the two panels in Fig.\ref{fig:pot_polym}  shows a significant decrease in the electric potential for almost linear configurations. 

It can also be appreciated through at the right end of the polymer a dip in the potential is visible, mirrored by a similar dip near the pore exit. The longitudinal contour in the top panel of  Fig.\ref{fig:pot_polym} is taken exactly at the time when the polymer is at mid-point during the translocation event. The contour shows that the part of the polymer external to the pore is only partially screened, and explains the drop of the potential to negative values (bottom panel). 

Note, that the polymer carries additional charge from the counterions while translocating, charges that mostly contribute to the total ionic current. Near the pore entrance and exit, parts of the polymer are left outside of the pore and in these regions, the polymeric charges are only partially screened by the counterions. The observed dips appear in proximity of the pore  entrance and exit since the unbalancing between the polymer beads and the counterions are caused by the attraction of latter within the pore region from the pore surface charge. Such attraction prevails over the strong repulsion between like charges under confinement. Due to the partial screening of the polymer in these regions, the polymeric region is stiffer than it is in bulk conditions, creating a sort of extended confinement effect.

In order to further elucidate the differences of the open and filled pores, we turn our attention to the conductance across the respective nanopores for different concentrations of the electrolytic solution. The results are summarized in Fig.\ref{fig:conduct} and reveal a linear variation with the applied potential difference for both the open and the filled pores. First, the importance of the salt concentration is clearly visible. A small increase in the conductance is seen for the filled pore as compared to the open pore. This observation can be attributed to the fact that the polymer is charged, thus the interaction with the surrounding ions can be stronger allowing the polymer to drag the coions, that is additional charge,  with it during the translocation. In addition, for the open pore, the conductance increases for higher salt concentrations. This is intuitive, as more ions are  added to the solution. For the closed pore, again the conductance increases in a good comparison to relevant experimental data \cite{kowalczyk2012measurement}.

In order to analyze the  variation for a specific applied potential difference, we  focus on the relative conductance in Fig.\ref{fig:relconduct}. For  the same potential difference, an increase of the salt concentration leads to a decrease in the relative conductance. This was evident for all potential differences in our work and agrees with recent experimental findings
\cite{smeets2006salt,kesselheim2014origin}.  Note, that according to the experimental data, this behavior  should be reversed  for a concentration over 300mM, but  such salt concentrations were not considered here. Note, that our results show a  highly non-linear behavior versus the potential difference for $e\Delta V/k_B T > 1$, but a more thorough investigation is left for future work.

\begin{figure}[h]
\centering
\includegraphics[width=0.4\linewidth]{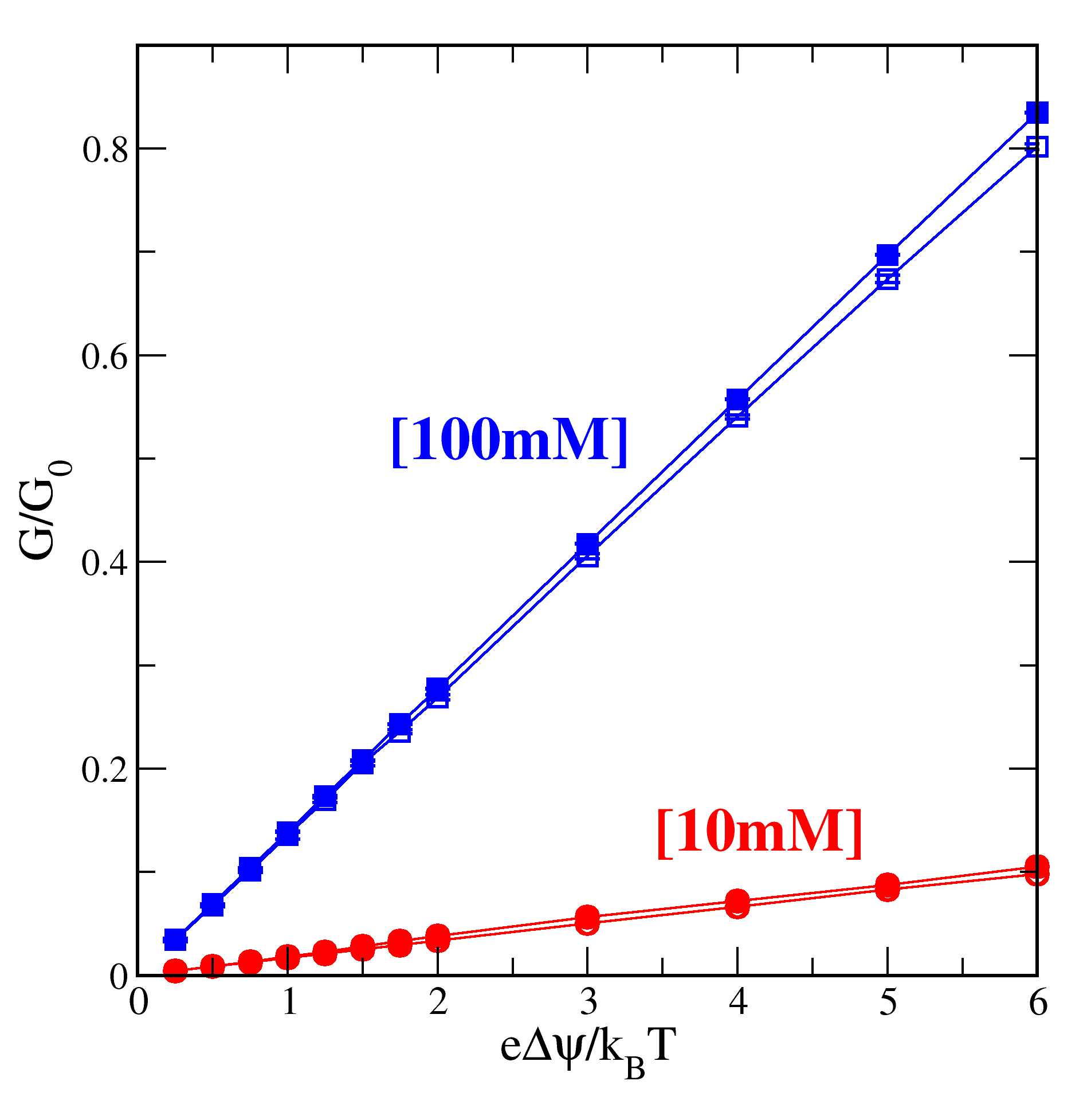}
  \caption{Conductance $G/G_0$ vs the applied tension for open pores (open symbols) and closed pores (filled symbols). The reference conductivity is $G_0=e^2/m v_T L_p$, where $v_T$ is the thermal velocity and $L_p$ is the pore length, and for the present pore equals 
  $G_0 = 75.3 ~ mS$. Results are shown for two concentrations, 10mM (red) and 100mM (blue). 
 }
  \label{fig:conduct}
\end{figure}

\begin{figure}[h]
\centering
\includegraphics[width=0.4\linewidth]{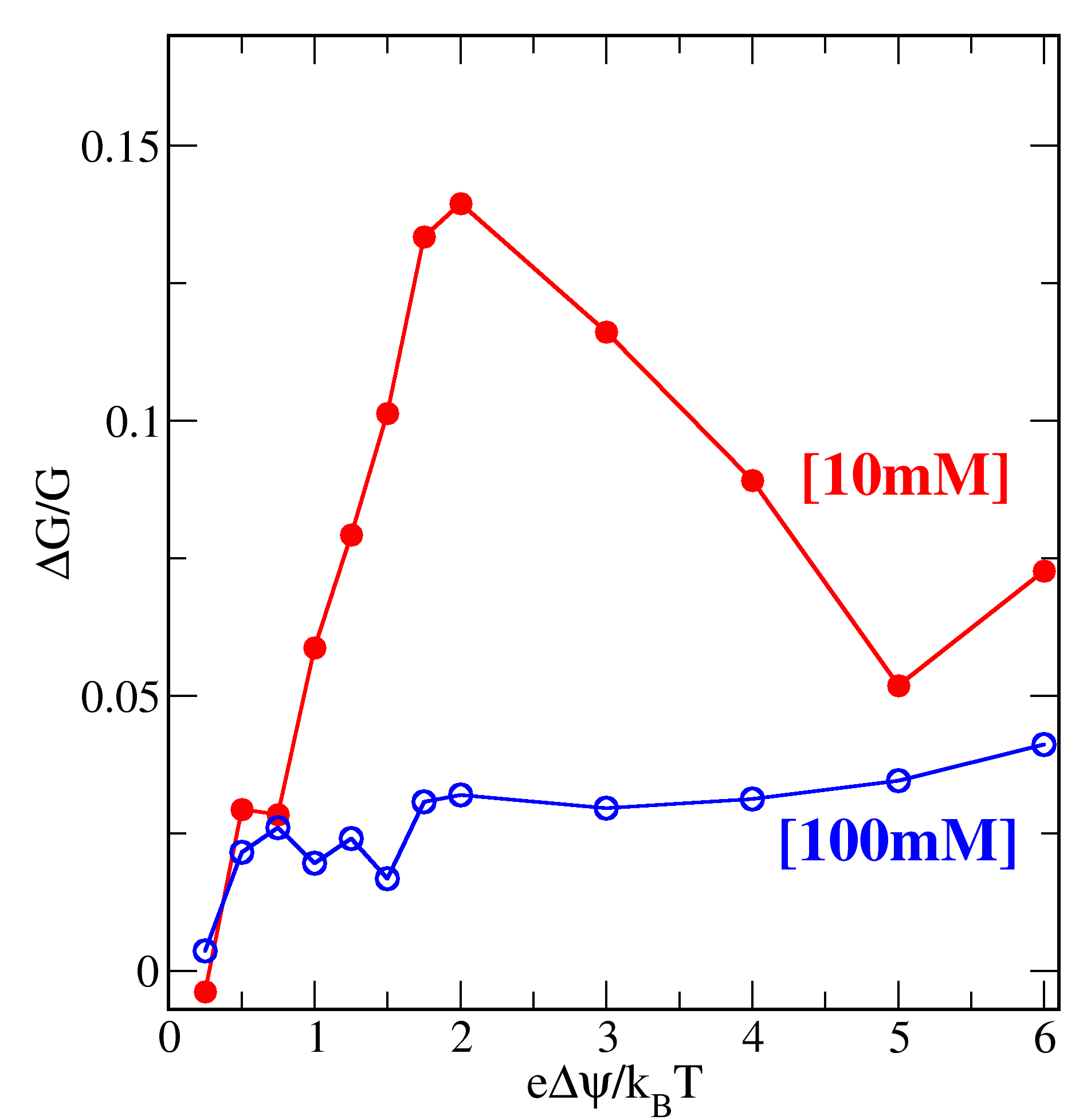}
  \caption{The variation of the relative conductance ($\Delta$G/G) with the potential difference applied at the pore is shown for open pores (open symbols), as well as for the pores being translocated by a polymer (filled symbols). Results are shown for two concentrations, 10mM (red) and 100mM (blue). 
  }
  \label{fig:relconduct}
\end{figure}







\subsection{Polymer dynamics}

As a final analysis, we focus on the dynamics of the translocating polymer.
During a translocation event, the polymer alters the pore conditions and modifies the passing ionic current,
giving rise to detectable current blockades \cite{kasianowicz1996characterization}. These blockades are indicative of how much ionic current is being screened off from the pore during a translocation event and is mainly based on excluded volume effects. 
The  fluctuations in the ionic blockades are directly related to the dynamics of the polymer conformations in the pore and also denote the start and the end of a translocation event.

\begin{figure}[h]
\centering
\includegraphics[width=0.4\linewidth]{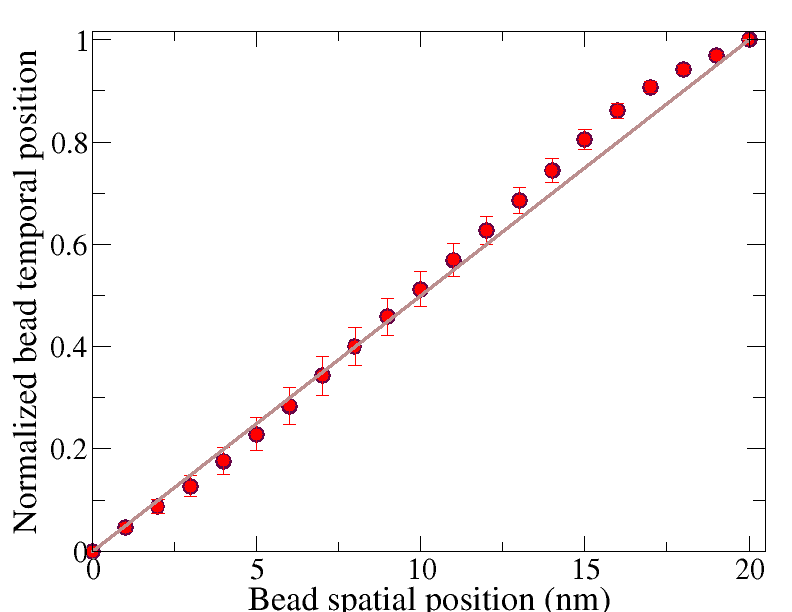}
  \caption{The relation between the spatial and temporal position of the polymer beads is shown. A deviation from the constant velocity assumption (brown line) can be seen. The error bars denote the uncertainty in the measurements. The results are shown for a 0.1M salt solution.}
  \label{fig:position}
\end{figure}

We monitor the correlation between the temporal and the spatial position of the polymer during a translocation event. The results for a 100mM solution are normalized with respect to the total time of the event and shown in Fig.\ref{fig:position}. We observe a deviation from the linear behavior, especially after the beginning and towards the end of the event (a linear behavior corresponds to a constant velocity assumption throughout the process).  Accordingly, at those regions, the translocation velocity is not constant, that is, each bead does not pass a certain point (for our analysis the midpoint of the pore) at regular time intervals. In this respect, from the current measurements the exact start and end of the translocation event cannot be exactly set: at the beginning of the process, the results lie below the constant velocity assumption denoting that the translocation takes more time than expected; for the final part of the process, the results lie above the constant velocity assumption and the translocation is faster than expected. 

The standard deviation in these measurements is also shown and shows a larger distribution of values at the midpoint of the translocation event. The start and the end of each event is taken from the ionic current blockades and those points are known with certainty. In this respect, the uncertainty of the first and final beads are zero, but increase towards the mid-point of the process. Our results compare well with recent experimental observations, which have reported an increase in the translocation velocity toward the end of the process \cite{plesa2014velocity}. In the experimental work, though, the real results are always above the constant velocity assumption. This difference can be  ascribed to the lower statistics and much shorter polymer lengths in our work as compared to the experiments. Increasing both aspects could shift our results more closer to the experimental ones. However, both  experimental and our work confirm the fact that the speed-up of the translocation at its end phase can be attributed to conformational rearrangements of the polymer, such as unfolding and refolding as translocation initiates or completes, respectively.


\section{Summary\label{sec:summ}}

This paper presents a multiscale multiphysics approach able to deal with different descriptions of matter. The scheme can efficiently describe charged solutes moving in a charged solution. It couples numerically an implicit electrokinetic solver to the Molecular Dynamics method. The main ingredients of the approach, as well as the coupling procedure have been extensively presented. 

Within our computational scheme, a continuum description is taken for the solvent made of a ternary charged mixture. The method takes into account electrostatic and viscous interactions, as well as the transport of different solvent species. The dynamics of the solute particles are described through Langevin equations with forces including solute-solute, solvent-solute, and electrostatic terms. For the former bonding and excluded volume terms are taken into account. The Eulerian and Lagrangian frameworks are seamlessly coupled through the drag force guiding the solvent-solute interactions.

In order to validate our numerical scheme we focused on the electrophoretic flow of charged polymers. A very good comparison with available experimental data was reached for a number of polymer lengths and salt concentrations.
As a first real application of our multiscale method, we have chosen the polymer translocation through very narrow pores. This process can strongly manifest the validity of our approach, as it includes all the necessary ingredients of the method and can be compared against a large number of relevant experimental and theoretical studies. Different kinds of translocation events have been observed, similarly to relevant experimental findings. Our approach offers insight into these events, through an analysis of the electric field in and around the nanopore, the conductance across the nanopore, the current blockades signature, the distribution of events and translocation times, etc. 

In the end, the newly developed method has proven efficient to deal with the motion of charged objects within a charged environment in complex geometries. The validation and first application shows a stable and consistent behavior, together with very reasonable agreement to available experimental observations. One aspect that will be important to explore in the future is to utilize monomeric charges close to that of DNA, that is, two electron charges per base pair. To this end, the propagation of the LB-MD scheme will need to employ specific propagators to handle stiff forces, such as the approach proposed in \cite{SCHILLER20142586}. 

The impact of our multiscale methodology in modeling complex phenomena occurring in a fluidic environment is potentially high. The drawbacks based on the necessary coarse-graining of the solute, the discrete description of the solvent and the implicit nature of the ions is overcome by the ability to stretch the temporal and spatial scales of the simulated systems. In the end, this scheme can be applied to a wide range of phenomena in nanofluidics, biophysics, polymer science, etc. Overall, the method is very flexible and capable to stretching the spatial and temporal scales in order to realize very large scale simulations. It can also deal with different time scales for the different components (fluid and molecule) allowing the modeling of problems influenced by details at both fine and coarse scales. The scheme can easily and efficiently be applied in the low salinity regime and include chemical specificity. In fact, we have already implemented potentials for biomolecules, such as DNA and RNA, and are currently extending this work in dealing  with processes involving biomolecules in electrolytic solutions.

\section{Acknowledgments}
The authors wish to thank C.Holm, G. Rempfer, and F. Weik for useful discussions and suggestions. Funding from the German Funding Agency (Deutsche Forschungsgemeinschaft - DFG) within the framework of the SFB716  collaborative network "Dynamic simulation of systems with large particle numbers" is greatly acknowledged. 
 MF acknowledges support from the Juniorprofessorenprogramm funded by the Ministry of Science, Research and the Arts Baden-W\"urttemberg (MWK).





\bibliography{transloc_LBMDelectrokin_meth}{99}

\end{document}